# Spectral tuning of multimode three-dimensional photonic crystal cavities for enhanced anti-Stokes Raman scattering

**Jing Ouyang and Minghao Qi[*]**

*School of Electrical and Computer Engineering, and Birck Nanotechnology Center*

*Purdue University, West Lafayette, IN 47907, USA*

[*]*Corresponding author: mqi@purdue.edu*

Multimode hollow microcavities in three-dimensional (3D) photonic crystals (PhCs) are designed for achieving enhanced coherent anti-Stokes Raman scattering, which requires a cavity to have three high quality-factor (Q) modes with equally spaced resonant frequencies. Cavities in 3D PhCs allows more flexibility in design and tuning than their 2D slab counterparts, since radiation loss that degrades Q can be suppressed by the 3D photonic band gap. We first tune all the mode frequencies simultaneously by changing the material and geometry of the cavity based on perturbation theory. Spectral spacings between the multiple modes are adjusted according to the symmetry, volume and field distribution of their mode profiles. The frequency and field distribution of the resonant modes are computed by solving Maxwell's equations in the frequency domain.

Photonic crystal (PhC) cavities with mode tunability have been explored for many applications such as microcavity sensors[1, 2], and enhanced Raman amplification[3]. These applications exploit the evolution of specific modes of a PhC cavity as the geometry, material or environment of the cavity changes. While mode tuning in two-dimensional (2D) slab PCs has already led to useful applications, the tuning is restricted to only one or two modes[1-3]. Additionally, the spectral tuning of 2D slab PhC cavities also affects their quality factors (Qs), and in most cases, the spectral tuning and high-Q cannot be simultaneously achieved. The ideal platform for spectral tuning is therefore a 3D PhC with a large, omnidirectional photonic band gap (PBG)[4], which allows a large range of spectral tuning without the degradation of Qs. Moreover, the unique ability to strongly confine the light in air makes 3D PhC cavities attractive to gas phase sensing applications, through enhanced coherent anti-Stokes Raman scattering (CARS)[5].

The intensity of CARS is proportional to the square of the pump optical power and to the Stokes pump power. 3D PhC cavities have unique capability of confining light in air and could significantly enhance the CARS if a single cavity can support three modes at frequencies $\omega_{pump}$, $\omega_{Stokes}$, and $\omega_{anit-Stokes}$; and the following is satisfied:

$$\omega_{pump} - \omega_{Stokes} = \omega_{anti-Stokes} - \omega_{pump} \tag{1}$$

Moreover, $\omega_{pump}$ should ideally be at the center of the PBG, so that the largest possible enhancement of optical power can be achieved for a given size of PhC. Additionally, the Stokes and anti-Stokes modes can also be close to the center of the PBG, thus improving their optical confinement. Furthermore, to detect a specific Raman transition of a molecule, the frequency spacing, $\omega_{pump} - \omega_{Stokes}$, must be independently tunable. In this letter, we present a perturbation-based method to tune the resonant frequencies of multiple modes associated with a single "hollow" cavity in a 3D PhC (Fig. 1a and 1b).

Two types of 3D PCs[6, 7] with PBGs in the near infrared were successfully fabricated via the layer-by-layer approach, which allows the introduction of microcavities with high accuracy. In particular, one type of the above bears striking similarity to 2D PC slabs[7, 8]. This type of 3D PC is an *fcc* lattice of air cylinders in dielectric (lattice constant *a*, $\varepsilon=12$, radius $0.293a$, height $0.93a$), oriented along the {111} direction[9]. The 3D structure is an alternating stack of the two basic 2D photonic crystal slab geometries (Fig. 1a and 1b): dielectric rods in air (rod layer) and air cylinders in a dielectric slab (hole layer)[9]. The 3D PC which was fabricated by over-etching method[7] has sculptured hexagonal rods in rods-in-air layers. Here, we use 3D photonic crystal with circular rods (radius $0.115a$, shown in Fig. 1a and 1b) which has a complete PBG from 0.509 to 0.638 ($2\pi c/a$), or 22% of the center frequency of the PBG.

Acceptor-type defects created by removing dielectric material have cavity modes that are pulled up from the bottom edge of the band gap and less strongly localized in the dielectric material than those of donor-type defects[10]. Multimode point-defect cavities by adding dielectric material classify their modes by point group symmetry, and normally have close-spaced resonant frequencies if not degenerate[11]. Our initial multimode cavity is an L3 defect, formed by removing three consecutive rods along the *x* direction in one of the rod layers (Fig. 1a), creating three modes with $\omega_{lower}$, $\omega_{middle}$ and $\omega_{upper}$ (shown in Fig. 1c-1e) equal to 0.544, 0.573, 0.61 ($2\pi c/a$), respectively. Two of the modes are even and one is odd in $\vec{E}_z$ with respect to the center of the cavity. The symmetry is later used for mode frequency tuning to achieve equal and variable spectral spacing. The modes are computed by solving Maxwell's equations in the frequency domain[12], with a resolution of 25 pixels/$a$ (about 18 pixels per in-plane period of the $x$-$y$ 2D cross section) and a supercell of 8, 6 and 6 periods in the $x, y, z$ directions, respectively. The cross section of one mode in $y$-$z$ plane is shown in Fig. 1b.

Direct analysis can be applied to explain the resonance frequency shift of PC cavities due to fabrication disorder[13]. Likewise, known cavity modes like those shown in Fig. 1 can be tuned for desirable effects by applying perturbation to the system, where the first-order perturbation theory for Maxwell's equations can be applied[14]. If $\vec{E}$ is approximated as continuously varying, the small change $\Delta\omega$ is given by the first-order perturbation,

$$\Delta\omega \approx -\frac{\omega}{2}\frac{\int \Delta\varepsilon |\vec{E}|^2}{\int \varepsilon |\vec{E}|^2} \qquad (2)$$

where $\vec{E}$ denotes the electric field of resonant mode at frequency ω of the PC cavity. For our designed cavity, one can add three rods of low dielectric constant material, such as $SiO_2$, back to where three original rods were removed. Due to the considerably smaller refractive index of $SiO_2$ ($n$=1.5, where $n$ is refractive index) in comparison to the rest of the dielectric material in a 3D PC, typically Si ($n$=3.4), adding small $SiO_2$ rods will not significantly alter the original mode profiles, and Eq. (2) indicates that the resonant frequencies of the cavity modes will decrease.

We first aim to move the middle of the three resonance frequencies to the center of the PBG. To show how the modes evolve, we add three identical $SiO_2$ rods to where the three original Si rods were removed (Fig. 2 inset). By increasing the radius of the three $SiO_2$ rods simultaneously, the resonance frequencies of all three modes are pulled down, an effect that can be explained with the first-order perturbation theory. For the modes shown in Fig. 1c-1e, the middle mode is already located very close to the gap center. Meanwhile, when the middle mode is located above the gap center, this step is necessary (*i.e.* when four surrounding rods have small radius, discussed later in Fig. 4).

The second goal is to adjust the spacing between the three modes to achieve:

$$\omega_{midde} - \omega_{lower} = \omega_{upper} - \omega_{midde} \qquad (3)$$

We exploit the fact that the middle mode is an odd mode with respect to the cavity center, thus the optical field is around zero near the cavity center. Therefore if one selectively increases the radius of the $SiO_2$ rod at the center of the cavity, the resonance frequency of the middle cavity won't be affected appreciably, while the lower and upper modes, being even modes, will have gradually decreased resonant frequencies (Fig. 3). Effectively, $\omega_{upper} - \omega_{middle}$ decreases, $\omega_{middle} - \omega_{lower}$ increases. Fig. 3 indicates that for $r=0.08a$, the equal spacing condition is met for the three modes.

The third objective is to control the magnitude of the frequency spacing. This is important for practical applications such as gas phase sensing, as the cavity mode spacing must match a characteristic Raman transition of a molecule to be detected, such as Sarin or other chemical warfare agents. We choose to adjust the effective size of the cavity by changing the radius of the four rods surrounding the cavity (the dark brown rods in Fig. 1a). Intuitively, a larger-sized cavity tends to accommodate more modes with smaller mode frequency spacing. This is confirmed in Fig. 4: when the radius of those rods decreases, the mode separation also decreases.

Another important requirement for CARS is phase matching, as CARS is a coherent optical process. For modes in microcavities, phase matching is the overlap between the three modes (Stokes, pump and anti-Stokes). At first glance, since the three modes shown in Fig. 1c-1e have different symmetry with regard to the center of the cavity (two even and one odd), one would expect that the mode overlap between them would be very small. However, due to the coherent nature of CARS, the mode overlap $y$ is calculated as:

$$y = \frac{\int (\vec{E}_p \cdot \vec{E}_s^*)(\vec{E}_p \cdot \vec{E}_{as}^*) dV^2}{\int |\vec{E}_p|^2 |\vec{E}_s|^2 dV \cdot \int |\vec{E}_{as}|^2 |\vec{E}_p|^2 dV} \qquad (4)$$

where $\vec{E}_s$, $\vec{E}_p$, and $\vec{E}_{as}$ are the electric fields of the Stokes, pump, and anti-Stokes modes. If we assume these modes correspond to the three modes of an L3 cavity as shown in Fig. 1c-1e, respectively, the appearance of $\vec{E}_p$ twice in the numerator ensures that its odd symmetry will not cancel out $\vec{E}_s$ and $\vec{E}_{as}$, which both have even symmetry. Numeric calculation shows a mode overlap around 0.52, while not close to 1, is nevertheless not a limiting factor for our CARS enhancement.

Finally, we discuss the potential enhancement factor. A cavity can enhance the resonance optical power roughly by a factor of $Q/V_{eff}$, where $V_{eff}$ is the effective volume of the cavity mode. Since our designed cavity resonates simultaneously with pump frequencies $\omega_{pump}$, $\omega_{Stokes}$, and Raman emission frequency $\omega_{anit-Stokes}$, then the emission intensity could be enhanced by a factor roughly proportional to $Q_{pump}^2 Q_{Stokes} Q_{anti-Stokes}/V_{eff}^4$. If the $Q/V_{eff}$ can all reach the range of $10^4$, one would achieve an enhancement of CARS by 16 orders of magnitude, roughly equivalent to any other known enhancement techniques. Even for a 3D PhC with limited number of layers, *i.e.* 13 layers (roughly 2 periods in the vertical direction), the total enhancement factor could reach $10^{12}$, due to the small mode volumes of those modes. Fig. 1b shows that 4 layers had been fabricated in silicon with the cavity embedded in the 4$^{th}$ layer.

In conclusion, we presented a systematic approach to independently tune three modes of a 3D photonic crystal cavity. By carefully examining the field distribution of each mode and introducing perturbation into the system, we fix the resonance frequency of one mode while moving those of the other two modes to desirable positions. Such conditions are necessary for applications involving multiple modes, *e.g*, enhanced CARS in a single 3D PC cavity. The ultra-small footprints of 3D PC cavities allows a large quantity of cavities to be integrated in a chip-based platform, with each cavity tuned to enhance a characteristic Raman transition of a gas-phase molecule, thus achieving chemical sensing. Moreover, the CARS could also be viewed as a special case of four-wave mixing (FWM), and frequency upconversion with an efficiency approaching the quantum limit (50%) has been demonstrated[15]. Thus 3D PhC cavities filled with highly non-linear materials could offer high-efficiency FWM at low pump powers.

We acknowledge the support from the Defense Threat Reduction Agency (DTRA) under contract HDTRA1-07-C-0042. Computational resources are supported by the Network for Computational Nanotechnology (NCN) and the Rosen Center for Advanced Computing (RCAC) at Purdue.

**Figure Captions**

Fig. 1. (a) Schematic of the multimode cavity in a 3D PC, which consists of two alternating layers: one rod layer (in yellow) and one hole layer (in blue). The cavity is formed by modifying the diameters of seven rods which form a hexagon (two vertices in blue and four in dark brown) with a center rod (in red). Different colors indicate their different functionality in mode tuning. (b) Scanning electron micrograph of a 4-layer 3D PhC made from silicon, with a cavity formed by removing two adjacent rods (encircled). (c)-(e) Horizontal cross sections of the $E_z$ profiles of the three modes of an L3 cavity, where the red and blue rods are completely removed, and the dark brown rods have the same diameter as the regular rods.

Fig. 2. Tuning of resonance frequencies by adding three rods of low refractive index (shown in inset). The square, rhombus and circle marked lines correspond to the upper, middle and lower modes, respectively. The dashed line denotes the center of the PBG. Shaded areas indicate the edges of the PBG.

Fig. 3. Normalized resonance modes as a function of the radius of the center rod. Resonance frequencies of upper mode (square line) and lower mode (circle line) gradually decrease while that of middle mode (rhombus line) remains relatively unafftected. The equal spectral spacing of cavity modes is met when $r=0.08a$, where $r$ is the radius of center low-index rod.

Fig. 4. The resonance frequencies of cavity modes change as a function of the radius of the four surrounding rods (shown in inset as bigger circles around the center of the cavity). The radius of the center low-index rod is fixed at 0.08a.

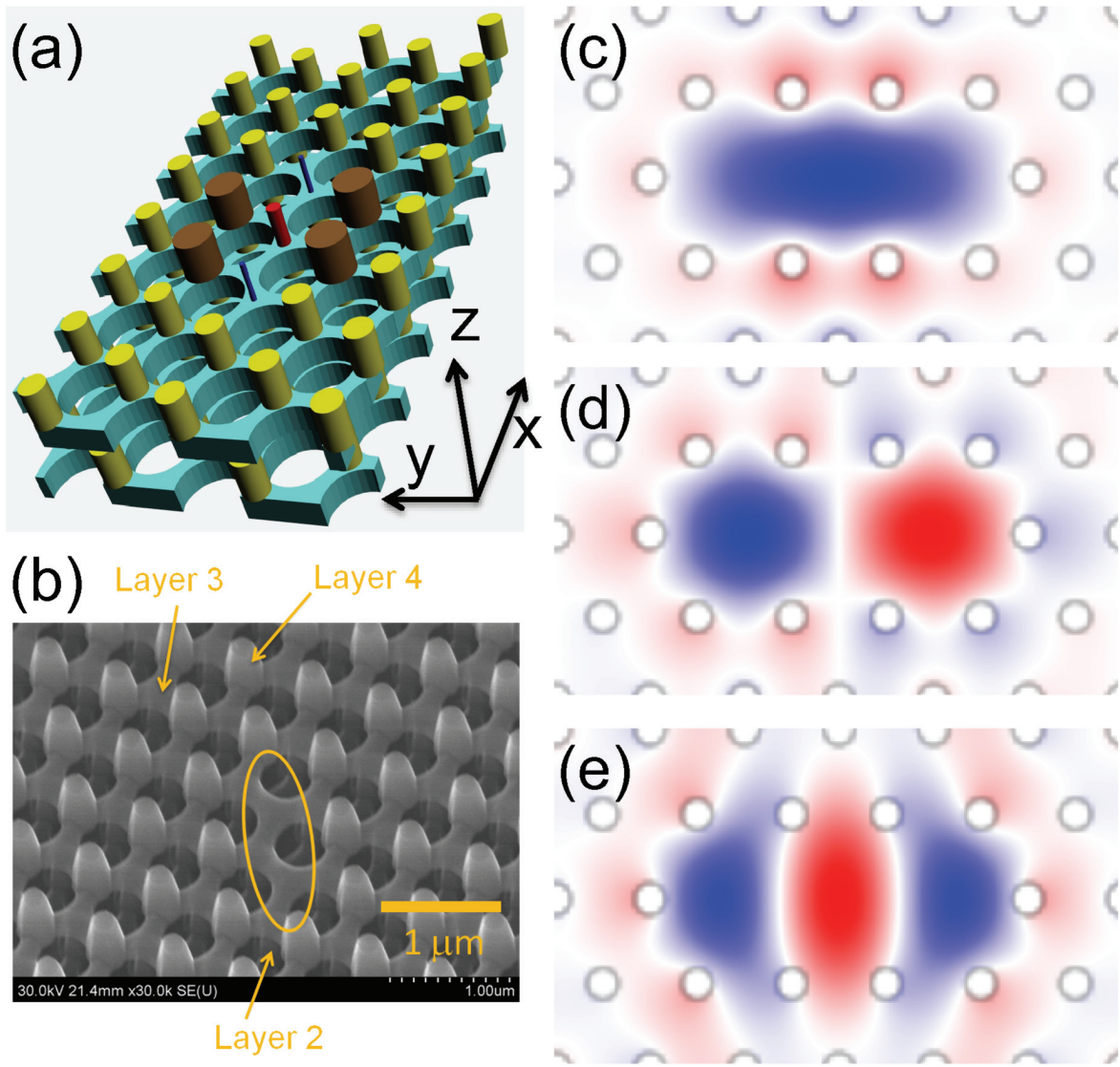

Fig. 1. (a) Schematic of the multimode cavity in a 3D PC, which consists of two alternating layers: one rod layer (in yellow) and one hole layer (in blue). The cavity is formed by modifying the diameters of seven rods which form a hexagon (two vertices in blue and four in dark brown) with a center rod (in red). Different colors indicate their different functionality in mode tuning. (b) Scanning electron micrograph of a 4-layer 3D PhC made from silicon, with a cavity formed by removing two adjacent rods (encircled). (c)-(e) Horizontal cross sections of the $E_z$ profiles of the three modes of an L3 cavity, where the red and blue rods are completely removed, and the dark brown rods have the same diameter as the regular rods.

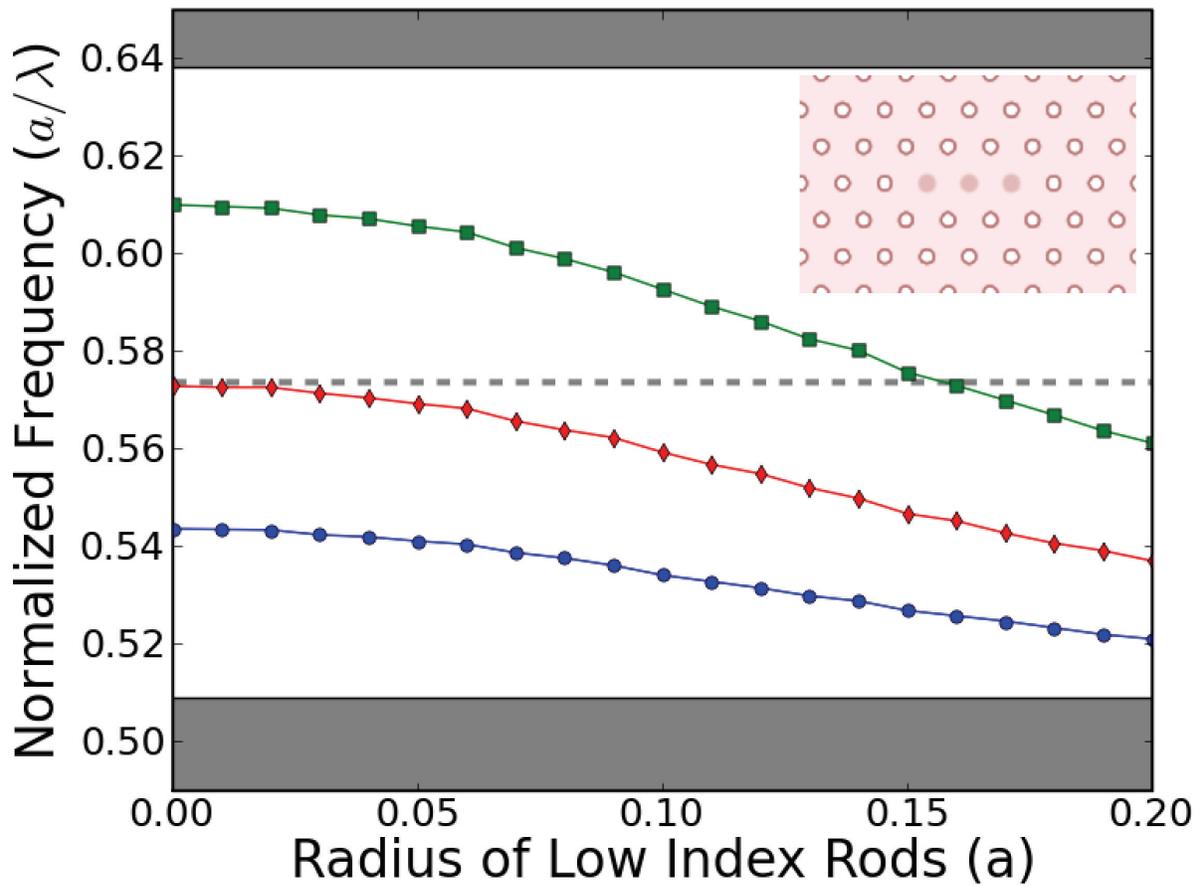

Fig. 2. Tuning of resonance frequencies by adding three rods of low refractive index (shown in inset). The square, rhombus and circle marked lines correspond to the upper, middle and lower modes, respectively. The dashed line denotes the center of the PBG. Shaded areas indicate the edges of the PBG.

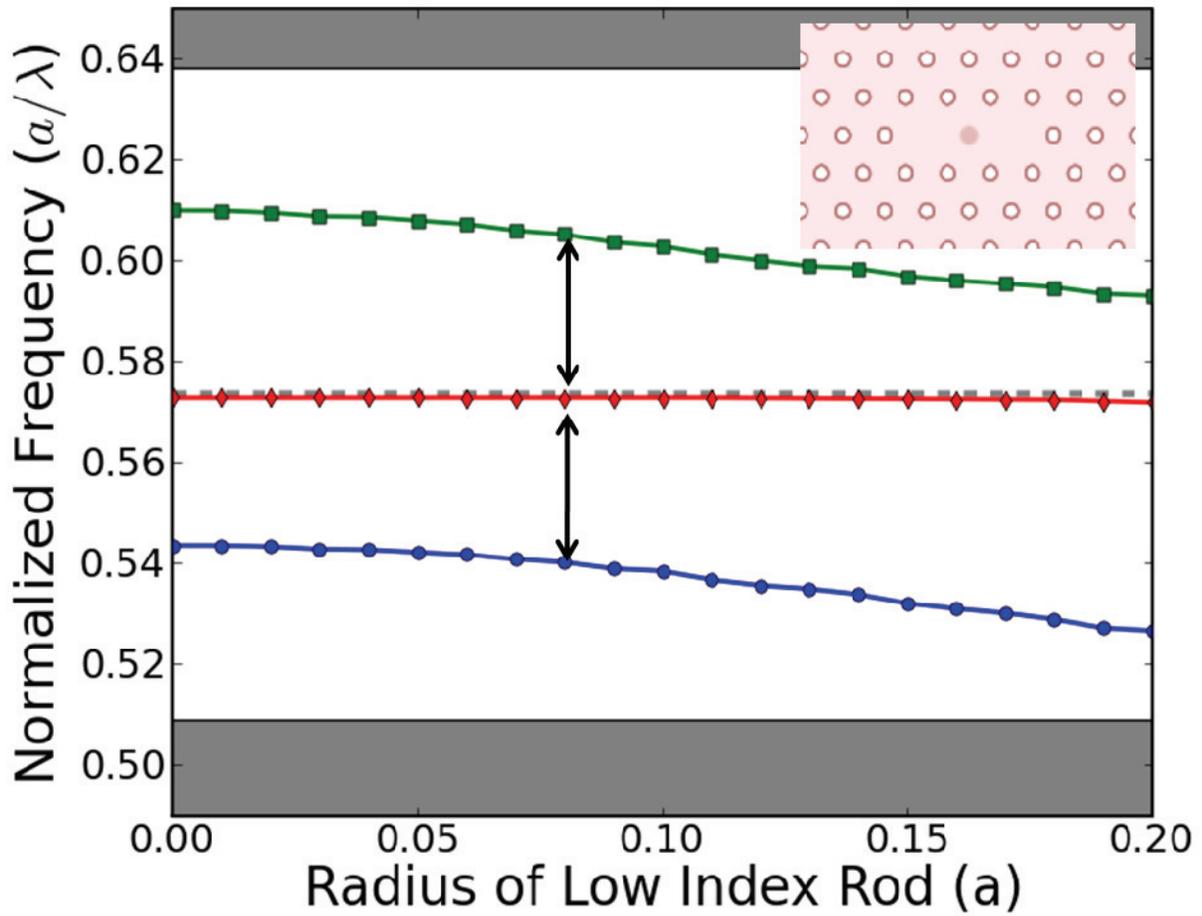

Fig. 3. Normalized resonance modes as a function of the radius of the center rod. Resonance frequencies of upper mode (square line) and lower mode (circle line) gradually decrease while that of middle mode (rhombus line) remains relatively unafftected. The equal spectral spacing of cavity modes is met when *r*=0.08*a*, where *r* is the radius of center low-index rod.

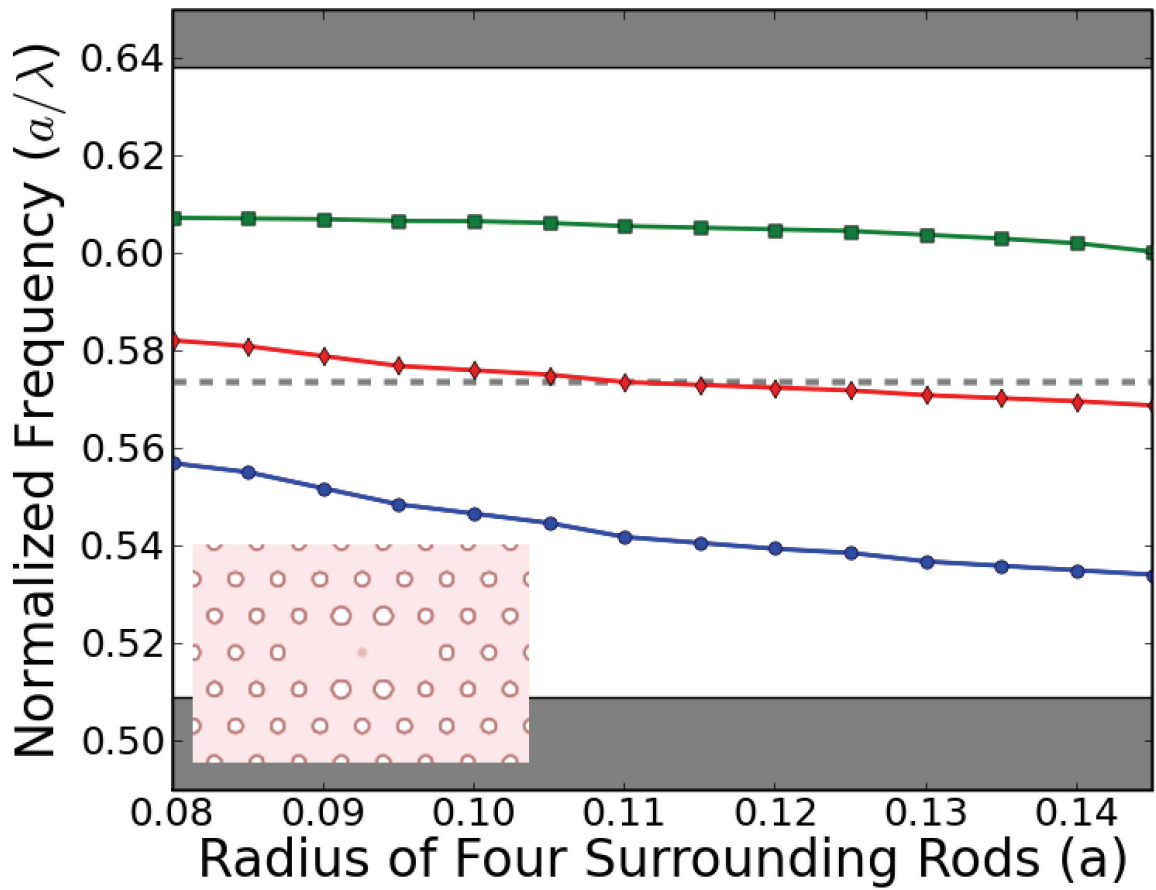

Fig. 4. The resonance frequencies of cavity modes change as a function of the radius of the four surrounding rods (shown in inset as bigger circles around the center of the cavity). The radius of the center low-index rod is fixed at 0.08a.